\documentclass[aps,twocolumn,nofootinbib,preprintnumbers]{revtex4}

\usepackage[dvips]{color}
\usepackage[normalem]{ulem}
\usepackage{amsmath}
\usepackage{enumerate}
\usepackage{amsfonts}
\usepackage{epsfig}

\newcommand{\be}{\begin{equation}}
\newcommand{\ee}{\end{equation}}
\newcommand{\bea}{\begin{eqnarray}}
\newcommand{\eea}{\end{eqnarray}}

\newcommand{\bln}{\begin{align}}
\newcommand{\eln}{\end{align}}
\newcommand{\bst}{\begin{split}}
\newcommand{\est}{\end{split}}
\newcommand{\bi}{\begin{itemize}}
\newcommand{\ei}{\end{itemize}}
\newcommand{\ben}{\begin{enumerate}}
\newcommand{\een}{\end{enumerate}}

\def\ov{\over}
\def\le{\left}
\def\ri{\right}
\def\ha{{1\over 2}}
\def\lam{{\lambda}}

\def\Sig{{\Sigma}}

\def\vev#1{\langle#1\rangle}

\def\th{{\theta}}

\def \th{{\theta}}

\def \lam {\lambda}
\def \om {\omega}

\def\sig{{\sigma}}
\def\ep{{\epsilon}}

\newcommand{\p}{\partial}

\newcommand\ga{{\ensuremath{{\gamma}}}}

\newcommand\De{{\Delta}}

\def\lam{{\lambda}}

\def\eeq{\end{equation}}

\newcommand\sF{{\ensuremath{{\mathcal F}}}}

\newcommand\sH{{\ensuremath{{\mathcal H}}}}

\newcommand\sN{{\ensuremath{{\mathcal N}}}}
\newcommand\sO{{\ensuremath{{\mathcal O}}}}

\newcommand\sJ{{\mathcal J}}
\newcommand\sT{{\mathcal T}}

\begin{document}

\title {Universality of the hydrodynamic limit in AdS/CFT and the membrane paradigm}

\preprint{MIT-CTP 3983}

\author{Nabil Iqbal and Hong Liu}

\affiliation{Center for Theoretical Physics,
Massachusetts
Institute of Technology,
Cambridge, MA 02139
}


\begin{abstract}
We show that at the level of linear response the low frequency limit of a strongly coupled field theory at finite temperature is determined by the horizon geometry of its gravity dual, i.e. by the ``membrane paradigm'' fluid of classical black hole mechanics. Thus generic boundary theory transport coefficients can be
expressed in terms of geometric quantities evaluated at the horizon. When applied to the stress tensor this gives a simple, general proof of the universality of the shear viscosity in terms of the universality of gravitational couplings, and when applied to a conserved current it gives a new general formula for the conductivity. Away from the low frequency limit the behavior
of the boundary theory fluid is no longer fully captured by the horizon fluid even within the derivative expansion; instead we find a nontrivial evolution from the horizon to the boundary. We derive flow equations governing this evolution and apply them to the simple examples of charge and momentum diffusion.
\end{abstract}

\maketitle
\newpage
\section{Introduction}
\label{intro}


The AdS/CFT correspondence \cite{AdS/CFT} is a powerful tool for understanding the dynamics of strongly coupled quantum field theories. Given the rapidly increasing number of models exhibiting such dualities, it would be desirable to be able to extract features
which are independent of the specific model.
Any interacting quantum field theory at finite temperature should be described
by hydrodynamics when viewed at sufficiently long length scales.
For those with a gravity dual, the bulk geometry involves a black hole with a non-degenerate horizon, and the UV/IR connection suggests that the field theory physics at long scales should be governed by the near-horizon portion of the dual geometry.
In fact, classical general relativity tells us that there is a precise sense in which any black hole has a fictitious fluid living on its horizon, in the so-called ``membrane paradigm''~\cite{thorne}.
It is thus tempting to identify the membrane paradigm fluid on the horizon with the low-energy description of the strongly coupled field theory. This connection was first made by~\cite{kss_stretch} and other related work  includes \cite{starmem,saremi,fujita}. See also~\cite{viscreview} for a general review of the hydrodynamic limit in AdS/CFT and related references on the subject.

In this paper we aim to clarify this connection by
comparing the linear response (to small external perturbations) of the horizon membrane fluid to that of the boundary theory fluid. Since in the hydrodynamic regime one is interested in
conserved quantities (or Goldstone modes), which in turn correspond to massless modes in the bulk, we will concentrate only on massless bulk modes in this paper. These cover
almost all interesting situations so far discussed in the literature. The only exception
is the bulk viscosity, as it cannot be associated with a massless degree of freedom in the bulk (see also comments in section~\ref{sec:concl}). We will leave discussion of the bulk viscosity and sound modes to a future publication.

We show here that that regardless of the specific model in question, the low-frequency limit\footnote{Here by low-frequency limit, we mean the lowest order term in the derivative expansions of frequency and spatial momenta.} of linear response of the boundary theory fluid is indeed completely captured by that of the horizon fluid.
In particular, this enables us to express a generic transport coefficient of the boundary theory solely in terms of geometric quantities evaluated at the event horizon of the black hole. For example, this gives a simple proof of the universality of the shear viscosity in terms of {\it the universality of the coupling of a transverse graviton}. We also give a new explicit expression for the conductivity of an arbitrary conserved current in the dual theory.

When moving away from the low frequency limit, however, the behavior
of the boundary fluid cannot be fully captured by the horizon fluid even within the
derivative expansion: even at generic frequency and momenta, the horizon response
always corresponds to that of the low frequency limit of the boundary theory.
Thus away from the low frequency limit, the full geometry of the spacetime plays a role.
To explore this we consider a fictitious membrane at each constant-radius hypersurface and introduce a linear response function for each of them. One can then derive a flow equation for the radius-dependent response function; at generic momenta this evolves nontrivially from the horizon to the boundary, where it determines the response of the dual field theory.
As an application of the flow equation we consider hydrodynamic diffusion. We give a simple derivation of the diffusion constants for charge and momentum diffusions and illustrate the difference between the diffusion phenomena observed at the horizon and at the boundary.

The plan of the paper is as follows. For the rest of this section, we introduce our conventions and notations for the gravity and field theory sides.
In section \ref{sec:mem} we give a quick review of the classical
black hole membrane paradigm. In section \ref{sec:linresp} we express linear response in AdS/CFT in a language similar to that of the membrane paradigm. Section \ref{sec:3} applies this language to the evaluation of zero-frequency transport coefficients in AdS/CFT. Sections \ref{sec:4} and \ref{sec:diff} are devoted to results at finite frequency, such as the flow equation and hydrodynamic diffusion. We conclude with a brief discussion in section \ref{sec:concl}.

\subsection{Gravity Setup}  \label{sec:1A}

On the gravity side, we will be examining very general black brane backgrounds, which we take to have the form
\be \label{bkG}
ds^2 = g_{rr} dr^2 + g_{\mu \nu} dx^\mu dx^\nu = - g_{tt} dt^2 + g_{rr} dr^2 +g_{ij} d x^i dx^j\ .
\ee
Indices \{$M$,$N$\} run over the full $d+1$-dimensional bulk, \{$\mu$,$\nu$\} over each $d$-dimensional constant-$r$ slice, and \{$i,j$\} over spatial
coordinates. We assume the above metric has an event horizon at $r = r_0$, where $g_{tt}$ has a first order zero and $g_{rr}$ has a first order pole. We assume that all other metric components are finite (i.e. neither zero nor infinite) at the horizon.

We take the boundary at $r=\infty$ and assume that the metric asymptotes to a structure that supports a gauge-gravity duality. 
We assume that all metric components and position-dependent couplings
depend on $r$ only so that we have translational invariance in $t$ and $x^i$ directions. We also assume the full rotational symmetry between $x^i$ directions, i.e.
\be \label{errk}
g_{ij} = g_{zz} \delta_{ij}
\ee
with $z$ one of the spatial direction.

Note that the metric (\ref{bkG}) does not have to be a spacetime metric, it could also be (for example) the induced metric on the worldvolume of a D-brane or a fundamental string in AdS.

We will often work in Fourier space on each constant-$r$ slice. For example for a scalar field $\phi$, we write
\be
\phi (r, x^\mu)= \int {d^d k \ov (2 \pi)^d} \, \phi (r, k_\mu) \, e^{i k_\mu x^\mu}, \qquad k_\mu = (-\om, \vec k) \ .
\ee
For simplicity of notation, we will distinguish $\phi (r, x^\mu)$ from its Fourier
transform $\phi (r, k_\mu)$ by its argument only.

\subsection{Field Theory Setup}

We will be relating these gravity backgrounds to quantum field theories taken at finite temperature. Consider a field theory containing an operator $\sO$ with an external classical source $\phi_0$. At the level of linear response theory the one-point function of $\sO$ is linear in $\phi_0$, and when expressed in Fourier space the proportionality constant is simply the thermal retarded correlator
$G^R$ of $\sO$
\be \label{defg}
\langle \sO(\omega,\vec{k}) \rangle_{\mathrm{QFT}} = -G^R(\omega, \vec{k})\phi_0(\omega, \vec{k})
\ee
where $\om$ and $\vec k$ denote the frequency and spatial momentum respectively (see e.g. \cite{kapusta}). The low frequency limit of this correlator is of physical
importance, as it defines a transport coefficient $\chi$:
\be \label{trco}
\chi = -\lim_{\om \to 0} \lim_{\vec k \to 0} {1 \ov \om} {\rm Im} G^R (\om, \vec k) \ .
\ee
Note that this definition essentially means that if we apply a time varying source $\phi_0(t)$, then in the low frequency limit the response of
the system is\footnote{Note that the real part of $G_R$ is even in $\om$ and we set the zero frequency part of $G_R$ to be zero since it gives rise to a contact term.}
\be \label{chidef2}
\langle \sO \rangle_{\mathrm{QFT}} = -\chi \partial_t \phi_0(t)
\ee
These transport coefficients are typically parameters in an effective low energy description
(such as hydrodynamics or Langevin equations) and once specified they completely determine the macroscopic behavior of the
medium.
A well-known example is the shear viscosity $\eta$, for which one takes $\sO = T_{xy}$, the off-diagonal component of the stress tensor. For DC conductivity $\sig$ one takes $\sO = J_z$, where $J_z$ is a component of the electric current. For quark diffusion constants characterizing the motion of a heavy quark moving in a quark-gluon plasma, $\sO$ is given by the forces acting on the quark~\cite{CasalderreySolana:2006rq}.

\section{The Classical Black Hole Membrane Paradigm} \label{sec:mem}

We begin our discussion with a brief review of the classical black hole membrane paradigm. Our treatment will not do sufficient justice to this elegant subject and will mostly
follow the formulation of the paradigm put forth in \cite{parikh}; for more detailed  exposition see~\cite{thorne}.


Imagine that we are observers hovering outside the horizon of a black hole. Since there is no (classical) way for the region inside the black hole to affect us, our effective action can be written as
\be
S_{\mathrm{eff}} = S_{\mathrm{out}} + S_{\mathrm{surf}}
\ee
where $S_{\mathrm{out}}$ involves an integration over the portion of spacetime outside the horizon and $S_{\mathrm{surf}}$ is a boundary term on the horizon, which can be determined by demanding that $S_\mathrm{eff}$ be stationary on a solution to the equations of motion.
Physically, $S_{\mathrm{surf}}$ represents the influence that the black hole horizon has on the external universe.
In practice, it is often more convenient to define $S_{\mathrm{surf}}$ on the ``stretched horizon'', which is a timelike surface of fixed $r$ just outside the
true horizon~(see \cite{pt} for extended discussion). This is also more concrete, since no observer can hover at the genuine horizon: the stretched horizon acts as a cutoff for the spacetime outside the black hole.

\subsection{The Membrane Conductivity}

Let us consider a bulk $U(1)$ gauge field with standard Maxwell action (see sec.~\ref{sec:1A} for our index conventions)
\be \label{Max}
S_\mathrm{out} = -\int_{r > r_0} d^{d+1}x \sqrt{-g} \frac{1}{4 g_{d+1}^2(r)} F_{MN} F^{MN} \ ,
\ee
where we have allowed a $r$-dependent gauge coupling $g_{d+1}$. The variation of this bulk action results in a boundary term at the horizon, which can only be canceled if
\be
S_{\mathrm{surf}} = \int_{\Sigma} d^d x \sqrt{-\gamma}\left(j^{\mu} \ov \sqrt{-\gamma}\right) A_{\mu},
\label{A_surfterm}
\ee
where $j^\mu$ is the conjugate momentum (with respect to $r$-foliation) of the field $A^\mu$
\be
j^{\mu} = -\frac{1}{g_{d+1}^2}\sqrt{-g}F^{r \mu}
\label{def_j}
\ee
and $\gamma_{\mu\nu}$ is the induced metric on the stretched horizon~$\Sig$.
Equation \eqref{A_surfterm} suggests that an observer hovering near the horizon will find that the horizon is carrying a membrane current
\begin{equation}
J_{\mathrm{mb}}^{\nu} \equiv \left(\frac{j^{\nu}(r_0)}{\sqrt{-\gamma}}\right) = -\frac{1}{g_{d+1}^2}\sqrt{g_{rr}}F^{r\nu}(r_0) \ .
\label{def_j_mb}
\end{equation}
Note that the ``Gauss's law'' that we obtain by treating $r$ as ``time'' becomes conservation of the currents $J^\mu_{\rm mb}$ and $j^\mu$ on any constant-$r$ slice.
\be
\p_\mu J^\mu_{\rm mb} = \p_\mu j^\mu = 0 \ .
\ee

While {\it a priori} the current $J^i_{\rm mb}$ (or $j^i$), which is determined by $F^{ir}$, and the electric field $E^i = F^{it}$ are independent variables, they
are in fact proportional to each other at the horizon. This can be seen as follows.
Since the horizon is a regular place for free in-falling observers, the electromagnetic
field observed by them must be regular. This implies that near the horizon,
$A^M$ can only depend on  $r$ and $t$ through their non-singular combination, the Eddington-Finklestein coordinate $v$ defined by
\be \label{EFC}
dv = dt +  \sqrt{g_{rr} \ov g_{tt}} {dr } \ .
\ee
This implies that
\be
\p_r A_i =  \sqrt{g_{rr} \ov g_{tt}} \p_t A_i \ , \quad r \to r_0
\ee
and with gauge choice $A_r =0$, we then have\footnote{This equation can be derived in a number of ways e.g. see \cite{parikh} for a gauge-invariant derivation.} (with $r \to r_0$)
\be
F_{ri} = \sqrt{g_{rr} \ov g_{tt}} F_{ti} \quad \to \quad
J^{i}_{\mathrm{mb}} = -\frac{1}{g_{d+1}^2}\sqrt{g^{tt}}F_t^i = \frac{1}{g_{d+1}^2}\hat{E}^i
\label{def_sigma_mb}
\end{equation}
Here $\hat{E}^{i}$ is 
an electric field measured in an orthonormal frame of a physical observer hovering
just outside of the black hole. From (\ref{def_sigma_mb}), it is natural to interpret $J^i_{\rm mb}$ as the {\it response} of the horizon membrane to the electric field $\hat E^i$, leading to a membrane conductivity
\be
\sigma_{\mathrm{mb}} = \frac{1}{g_{d+1}^2(r_0)} \ .
\ee
Note that (unlike those arising from conventional quantum field theories) this conductivity is frequency-independent and depends only on the gauge coupling at the horizon.

\subsection{The Scalar Membrane and the Shear Viscosity of the Membrane Paradigm Fluid}

Now consider a massless bulk scalar field with action
\begin{equation} \label{ree1}
S_{\rm out} = - \frac{1}{2} \int_{r > r_0} d^{d+1}x \sqrt{-g}\, {1 \ov q (r)} \, (\nabla\phi)^2,
\end{equation}
where $q(r)$ can be considered an effective ($r$-dependent) scalar coupling. The boundary term on the horizon resulting from variation of this action requires the addition of a surface action at the horizon
\be \label{selr}
S_{\mathrm{surf}} = \int_{\Sigma} d^{d}x \sqrt{-\gamma} \left(\frac{\Pi(r_0,x)}{\sqrt{-\gamma}}\right) \phi(r_0,x)
\ee
where $\Pi$ is again the momentum conjugate to $\phi$ with respect to a foliation in the $r$-direction,
\be \label{canM}
\Pi = -{\sqrt{-g} \ov q(r)} \, g^{rr}\partial_r\phi \ .
\ee

Following the discussion of an electromagnetic field, equation (\ref{selr}) now implies that
to an external observer the horizon appears to
have a ``membrane $\phi$-charge'' $\Pi_{\mathrm{mb}}$ given by
\be \label{memcharge}
\Pi_{\mathrm{mb}} \equiv \left({\Pi(r_0) \ov \sqrt{-\ga}}\right) = -{\sqrt{g^{rr}}\partial_r\phi(r_0) \ov q(r_0)}
\ee
Again, for a freely in-falling observer to find a nonsingular $\phi$,
near the horizon, $\phi$ should have the form $\phi (r,t,x_i) =
\phi (v,x_i)$, where $v$ is the Eddington-Finklestein
coordinate (\ref{EFC}). This implies $\partial_r\phi = \sqrt{{g_{rr} \ov g_{tt}}}\partial_t\phi$ and
\be
\Pi_{\mathrm{mb}} = -\frac{1}{q(r_0)}\sqrt{g^{tt}}\partial_t\phi(r_0) = -\frac{1}{q(r_0)}\partial_{\hat{t}}\phi(r_0)
\label{pimb_def}
\ee
where in the last equality we have passed to an orthonormal basis.
As in the electromagnetic case we can interpret $\Pi_{\mathrm{mb}}$ as the response of the horizon membrane induced by a local bulk field $\phi$
around the hole, leading to a membrane transport coefficient $\chi_{\mathrm{mb}}$ (compare with (\ref{chidef2}))
\be
\chi_{\mathrm{mb}} = \frac{1}{q(r_0)}
\label{def_chi_mb}
\ee
We emphasize that in deriving (\ref{pimb_def}) we did not take a low frequency or momentum limit; (\ref{def_chi_mb}) is the full response for generic momenta and frequencies.

We can now apply this discussion to the computation of the shear viscosity of the horizon
membrane by taking $\phi = h_{x}^y$, the off-diagonal component of the graviton.
We assume that there is no spatial momentum in $x-y$ directions and that the background matter stress tensor does not mix with $h_{x}^y$. Then the graviton is transverse and its action is simply that of (\ref{ree1}) with $q = 16 \pi G_N$ where
$G_N$ is the bulk Newton's constant. $\Pi_{\rm mb}$ can now be interpreted as $(T_\mathrm{mb})_y^x$, a component of the membrane stress tensor. From (\ref{pimb_def}) and (\ref{def_chi_mb}) we thus conclude
that~\cite{parikh,thorne}
\be \label{memD}
\eta_{\mathrm{mb}} = {1 \ov 16\pi G_N} , \quad \to \quad {\eta_{\mathrm{mb}} \ov s_{\mathrm{mb}}} = {1 \ov 4\pi}
\ee
where $s_{\mathrm{mb}} = 1/4G_N$ is (by definition) the entropy density per unit volume of the membrane fluid.

We see that the influence of the black hole on its surroundings can be taken into account by placing fictitious charges and currents on its horizon. Furthermore, these currents are related to applied
fields in a very simple way, fixed completely by the condition of horizon regularity. This leads to
simple expressions for transport coefficients such as $\eta_{\mathrm{mb}}$ and $\sigma_{\mathrm{mb}}$, although
it is not immediately clear whether these coefficients are in any way related to those that we calculate
from AdS/CFT.

\section{Linear response in AdS/CFT: Taking the membrane to the boundary} \label{sec:linresp}

Let us now turn to the corresponding problem in AdS/CFT. The massless bulk field $\phi$ with action \eqref{ree1} is now dual to an operator
$\sO$ in the boundary theory. Recall that in {\it Euclidean} signature, the relation between the dual theory generating functional and the on-shell supergravity action\footnote{In this paper we will only consider the gravity limit.} is given by
\be
\label{AdSident}
\left\langle \exp\left[-\int d^{d}x\;\phi_0 \sO\right] \right\rangle_{\mathrm{QFT}} = e^{-S_\mathrm{grav}[\phi(r\to\infty) = \phi_0]} \ .
\ee
In other words, we must find a classical solution for $\phi$ that is regular in the bulk and asymptotes to a given value $\phi_0$ at the boundary; derivatives of the on-shell gravity action with respect to this boundary value $\phi_0$ will give us correlators for $\sO$. Equation \eqref{AdSident} implies that the one-point function in the presence of source $\phi_0$ can be written as\footnote{Note that the limit on the right hand should be taken with some care. For example, for a massless field, one should take the part of $\Pi$ which goes to $O(1)$ at infinity.}
\be
\label{defexpO}
\langle \sO(x^{\mu}) \rangle_{\phi_0} = \lim_{r\to\infty}\Pi(r,x^{\mu})
\ee
where we have used the well known fact in classical mechanics that
the derivative of an on-shell action with respect to the boundary value of a field is simply equal to the canonical momentum conjugate to the field, evaluated at the boundary.

In Lorentzian signature the story is more intricate since one also needs to impose appropriate boundary conditions for $\phi$ at the horizon and
the analytic continuation of~\eqref{AdSident} will only yield Feynman functions.
A simple prescription for directly calculating boundary retarded two-point functions was given  in~\cite{realtime} and later confirmed in~\cite{Herzog:2002pc}.
Here we briefly summarize the prescription:
\ben
\item  Find a solution to the equations of motion that is in-falling at the horizon and asymptotes to a constant $\phi (r, k_\mu) \to \phi_0 (\om, \vec k)$ at the boundary $r  \to \infty$. 
\item To evaluate the on-shell action, we plug this solution into the action (\ref{ree1}) and integrate by parts. The action is reduced to surface terms at the boundary and horizon, which can be written
\be
\label{Sbd}
S = -\sum_{r = r_0,\infty}\frac{1}{2}\int\!{d^d k\ov (2\pi )^d}\,
\phi_0 (-k_\mu) {\cal F} (k_\mu,r) \phi_0 (k_\mu)
\ee
for some function $\sF$. The prescription is that the retarded Green's function is
\be \label{eo}
G_R (k_\mu) = \lim_{r\to \infty} {\cal F} (k_\mu,r) \ .
\ee
Writing $\phi(r,k_\mu) = f(r, k_\mu)\phi_0(k_\mu)$ so that $f(r,k)$ is normalized as $f(r\to\infty, k) = 1$, (\ref{eo}) can be written more explicitly as
\be \label{eor}
G_R (k_\mu) = \frac{1}{q(r)}\sqrt{-g}g^{rr} \partial_r f(r, k_\mu) \ .
\ee
\een
Note that the above prescription for calculating retarded two-point functions cannot be obtained from an action principle like~\eqref{AdSident}.

Our observation here is that the above prescription is in fact equivalent
to~\eqref{defexpO} at the linear level, now evaluated in Lorentzian signature, with the requirement that
$\phi$ satisfy in-falling boundary conditions at the black hole horizon.
To see this explicitly, taking a Fourier transform of~\eqref{defexpO} and comparing with \eqref{defg}, we obtain a simple formula for the thermal retarded correlator $G_R$:
\be \label{bdv}
G_R(k_\mu) = -\lim_{r \to \infty} {\Pi (r, k_\mu) \ov \phi (r, k_\mu)} \ .
\ee
Using \eqref{canM} one immediately sees that \eqref{bdv} is equivalent to \eqref{eor}.

We believe~\eqref{defexpO} and \eqref{bdv} provide a more fundamental prescription, as they are expressed in terms of quantities of clear geometric and physical meaning\footnote{For a massive field in AdS, \eqref{defexpO} and \eqref{bdv} become
$$
\vev{\sO(k)} =  \lim_{r \to \infty} r^{\De - d} \Pi (r), \quad G_R (k_a) = \lim_{r \to \infty} r^{\De - d} {\Pi (r) \ov \phi_0 (k_a)}
$$
where $\De$ is the dimension of the corresponding operator~$\sO$.}.
We note that with a different choice of boundary conditions at the horizon~\eqref{bdv} can also be used to calculate Feynman functions: this essentially follows from analytic continuation of \eqref{AdSident} to Lorentzian signature\footnote{An important subtlety here is that
the proper Euclidean continuation gives the so-called Hartle-Hawking vacuum and not the naive Schwarzschild vacuum.}.
It can also be immediately
generalized to fields of higher spin or fields with more general action. For a vector field, $\sO$ is replaced by a boundary current $\sJ^\mu$ and $\Pi$ by $j^\mu$ defined in (\ref{def_j}). For metric fluctuations $\sO$ is replaced by the boundary stress tensor $\mathcal{T}^{\mu\nu}$ and $\Pi$ by
\be \label{stens}
T^{\mu \nu} =  {\sqrt{-\gamma} \ov 16 \pi G_N} \le(K^{\mu \nu} - \gamma^{\mu \nu}
K^\lam_\lam \ri)
\ee
where $K_{\mu \nu}$ is the extrinsic curvature of a constant $r$-surface.

Comparing \eqref{defexpO} with \eqref{memcharge}, we see that \eqref{defexpO} expresses the AdS/CFT response in a language almost identical to that of the membrane paradigm, except that the ``membrane'' in question is no longer at the horizon but at the boundary $r \to \infty$. The object $\Pi$ which was loosely interpreted as a ``membrane response'' is now \emph{actually} the response of an operator $\sO$ in the dual theory; in other words, if we move the membrane from the horizon to the boundary, membrane paradigm quantities become concrete gauge theory observables. Furthermore, as can be checked explicitly, the in-falling boundary condition for $\phi$ at the horizon is precisely the regularity condition discussed for the membrane paradigm, i.e. $\phi$ can only depend on  $r$ and $t$ through the Eddington-Finklestein coordinate $v$~\eqref{EFC}. This is an expression of the physical statement that a local observer hovering at the horizon only sees things falling in, not coming out. An example illustrating this is given in Appendix \ref{sec:inf}.

\section{Low frequency limit} \label{sec:3}

In this section we first show that the field theory transport coefficient defined in (\ref{trco}) can be expressed solely in terms of quantities at the horizon. This  immediately gives a general proof of the universality of shear viscosity.
We then turn to a $U(1)$ vector field in the bulk and calculate the DC conductivity of the
corresponding conserved boundary current.

\subsection{General formula for transport coefficients}

Using (\ref{bdv}),  equation (\ref{trco}) can now be written as
\be \label{coe}
\chi = \lim_{k_\mu \to 0} \lim_{r \to \infty} {\Pi (r, k_\mu) \ov i \om \phi (r, k_\mu)} \ .
\ee
To compute (\ref{coe}) we write the equations of motion for
$\phi$ in a Hamiltonian form as
\bea \label{eom1}
\Pi & = & -{\sqrt{-g} \ov q(r)} \, g^{rr}\partial_r\phi \\
\p_r \Pi & = & {\sqrt{-g} \ov q(r)} \, g^{rr} g^{\mu \nu} k_\mu k_\nu \phi \ .
 \label{eom2}
\eea
Note that in the low frequency limit (i.e. $k_\mu \to 0$, with $\om \phi$ and $\Pi$ fixed), equations (\ref{eom1}) and (\ref{eom2}) become trivial
\begin{equation} \label{flow_scalar}
\partial_r \Pi = 0 + \mathcal{O}(k_\mu \om \phi) \qquad \partial_r(\om \phi) = 0 + \mathcal{O}(\omega\Pi) \ .
\end{equation}
Thus in the zero momentum limit the evolution in $r$ is completely trivial and \eqref{coe} can in fact be evaluated at any value of $r$! We will evaluate it the horizon where the in-falling boundary condition should be imposed.
As we noted as the end of section~\ref{sec:linresp}, the in-falling boundary condition for $\phi$ at the horizon is in fact equivalent to the condition of horizon regularity in the membrane paradigm, which from~\eqref{canM}--\eqref{pimb_def} gives,
\begin{equation} \label{pibc}
\Pi(r_0,k_\mu) =  {1 \ov q(r_0)} \sqrt{\frac{-g} {g_{rr}g_{tt}}} \biggr|_{r_0} i \om \phi(r_0,k_\mu) \ .
\end{equation}
We thus find the simple result
\be \label{ero}
\chi =  {1 \ov q(r_0)} \sqrt{\frac{-g} {g_{rr}g_{tt}}} \biggr|_{r_0} = {1 \ov q (r_0)} {A \ov V}
\ee
where $A$ is the area of the horizon and $V$ is the spatial volume of the boundary theory.
Given that entropy density $s$ of the boundary theory is given by $s = {A \ov 4 G_N V}$, $\chi$ can also be written as
\be \label{ratio}
{\chi \ov s} = {4  G_N \ov q (r_0)} \ .
\ee

We find that the ratio $\chi/s$ is given by the ratio of the Newton constant $G_N$, which characterizes the gravitational coupling, to the effective coupling $q(r_0)$ for $\phi$  at the horizon.
This is the result for the boundary fluid, but we see that it is closely related to the corresponding result for the horizon fluid, simply because the bulk evolution equations \eqref{eom1} and \eqref{eom2} are trivial in
the low frequency limit. Note that (provided that some form of gauge-gravity duality exists so that \eqref{defexpO} and \eqref{bdv} make sense) the precise asymptotic structure of the spacetime does not play an important role in our analysis.

Though we have used the example of a massless scalar field above, the discussion clearly applies to components of more general tensor fields. Our treatment can be applied to very general effective actions of the form
\be \label{moreG}
S = -\frac{1}{2}\int \frac{d\omega d^{d-1}k}{(2\pi)^d} dr\sqrt{-g}\left[\frac{g^{rr}(\partial_r\phi)^2}{Q(r; \omega,\vec k)} + P(r; \omega, \vec k)\phi^2\right],
\ee
provided that the flow equations \eqref{flow_scalar} remain trivial in the zero-momentum limit. This implies that $Q$ should go to a nonzero constant at zero momentum  and $P$ must be at least quadratic in momenta,
so a mass term would not be allowed. For (\ref{moreG}) the corresponding transport coefficient is given by
\be \label{ratio1}
{\chi \ov s} = {4  G_N \ov Q (r_0, k_\mu=0)} \ .
\ee
It is also straightforward to generalize the discussion to multiple coupled fields.

\subsection{Universality of Shear Viscosity}

The most obvious application of~\eqref{ero}--\eqref{ratio} is to the shear viscosity. For Einstein gravity coupled to matter fields, in the absence of a background off-diagonal component of the metric, the effective action for the transverse off-diagonal gravitons $h_{x}^y$, which is dual to $\mathcal{T}_{xy}$ in the boundary theory, is simply that of a massless scalar with effective coupling
\be \label{eei}
q(r) = 16 \pi G_N \ .
\ee
From (\ref{ratio}) we thus find the celebrated result
\be \label{etas}
{\eta \ov s} = {1 \ov 4 \pi}
\ee
Note that the universality of (\ref{etas}) can now be attributed to the universality of the effective coupling (\ref{eei}) for a graviton $h_{x}^y$.

The proof here is very general, applying to all Einstein gravity duals known so far, including
charged black holes dual to theories with chemical potential, the near-horizon
region of general Dp-branes, and recently discovered geometries dual to non-relativistic CFTs~\cite{nonrel1}.\footnote{An interesting example that violates our assumptions but still satisfies \eqref{etas} is the gravity dual to non-commutative $\sN = 4$ plasma studied in \cite{noncomm}. This system has broken rotational invariance and a boundary stress tensor that is dual not to the bulk graviton but to a linear combination of graviton and $B$ fields, and thus our analysis does not apply, though it would be interesting to see if it could be extended.}

It is also useful to compare this proof to earlier calculations. By considering diffusive shear modes on the stretched horizon \cite{kss_stretch}
derived a formula for the diffusion constant $\eta/(\ep + p)$ in terms of an integral from the horizon to the boundary, where $\ep$ and $p$ are energy and pressure density. \cite{buchel}
showed that $\eta/s$ is universal among a certain family of metrics by
reducing the integral to a total derivative and then evaluating it at the horizon.
A simpler proof along a similar line was given in~\cite{viscreview}. The proofs of~\cite{buchel,viscreview}
do not include theories with nonzero chemical potentials which have been checked separately \cite{chempot1,chempot2,chempot3,chempot4,chempot5}. Note that there is no conflict with
the integral formula of~\cite{kss_stretch} and our result (\ref{ratio}) since the diffusion constant $\eta/(\ep+p)$ involves a factor $\ep+p$ which generically cannot be expressed in terms of quantities at the horizon.
Our proof is closer in spirit to the discussions in \cite{blackholephysics,chempot1,universstress}, which use the Kubo formula and
are related to the universality of the absorption cross section of a minimally coupled scalar \cite{gibbons}. The treatment given here is more general and technically simpler. It also highlights the importance of the effective coupling at the horizon as the source of universality.

It has been conjectured~\cite{blackholephysics} that the value of $\eta/s$  in (\ref{etas}) is in fact a lower bound for all realistic matter.
Equation \eqref{ratio1} hints at how the bound could be violated. We need to find a theory whose ``effective'' gravitational coupling for the $h^{x}_y$ polarization at the horizon is {\it stronger} than
the universal value (\ref{eei}) for Einstein gravity. Gauss-Bonnet gravity as discussed in~\cite{Brigante:2007nu,Brigante:2008} (see also~\cite{Kats:2007mq}) is an example of this. There the effective action for $h_x^y$ has the form of (\ref{moreG}) with the effective coupling $Q(r)$ at the horizon satisfying (see 3.10 in \cite{Brigante:2007nu})\footnote{We note that while the formula \eqref{ratio1} does not immediately apply to more general higher-derivative gravity theories such as the $R^4$ theory studied in \cite{coupdep} due to the presence of terms in the action that are higher than quadratic in derivatives, it is conceivable that with some effort the discussion may generalized to include that case as well.} 
\be
\frac{1}{Q(r_0)} = \frac{(1-4\lambda_{\rm GB})}{16\pi G_N} \ .
\ee
Thus for $\lam_{\rm GB} > 0$ the graviton in this theory is more strongly coupled
than that of Einstein gravity and the value of $\eta/s$ dips below the value (\ref{etas}).
This also indicates that for $\eta/s$ to be arbitrarily small, the graviton has to be
strongly coupled, which was indeed observed in the Gauss-Bonnet example~\cite{Brigante:2007nu}.

Note that the formula (\ref{ratio}) for the specific case of shear viscosity in higher derivative gravity theories has been conjectured recently in~\cite{Brustein:2008cg}, where the authors also discuss the
importance of the strength of the coupling for $h_{x}^y$ and its relation to the violation of the viscosity bound.

\subsection{Gauge Fields and the DC Conductivity}

As another example of application of \eqref{ratio}, we now turn to the DC conductivity.
The bulk gauge field $A_M$ in AdS is now dual to a conserved current $\sJ^{\mu}$ in the boundary theory.
The action is given by \eqref{Max}, and the momentum
conjugate to this gauge field is given by (\ref{def_j}).
From (\ref{bdv}) and (\ref{defexpO}), the conductivity can be written as
\be
\langle \sJ^i (k_\mu) \rangle_\mathrm{QFT}  = j^i(r\rightarrow\infty) (k_\mu) \equiv \sigma^{ij} (k_\mu) F_{jt}(r \to \infty) \ .
\label{defsigma}
\ee
Equation \eqref{defsigma} defines AC conductivities which are related to the
retarded Green function of the boundary current $\sJ^i$ by
\be
\sigma^{ij}(k_\mu) = -{G_R^{ij}(k_\mu) \ov i\omega} \ .
\label{Asigma}
\ee
The DC conductivity is obtained by the zero momentum limit of the above equations.
Clearly at zero momentum due to rotational symmetry, $\sig_{ij} = \sig \delta_{ij}$ \footnote{Except in two spatial dimensions when we can also have an antisymmetric component proportional to $\epsilon^{ij}$.}.
Below $\sig$ without an explicit argument always refers to the DC conductivity.

The in-falling boundary condition at the horizon, which translates into  equation~\eqref{def_sigma_mb}, gives us
the ratio at the horizon :
\be
j^{i}(r_0) = \frac{1}{g_{d+1}^2}\sqrt{\frac{-g}{g_{rr}g_{tt}}}g^{zz}\biggr|_{r_0} F_{it}(r_0) \ .
\label{horizcond}
\ee
It can be readily checked that in the zero momentum limit the bulk Maxwell equations give (see e.g. \eqref{flow2} and \eqref{bianchi} of Appendix \ref{app:A})
\be
\p_r j^i = 0 + \sO(\omega F_{it}), \qquad \p_r F_{it} = 0 + \sO(\omega j^i)
\ee
implying that the relation \eqref{horizcond} actually holds for all $r$. Combining \eqref{defsigma} and \eqref{horizcond}, we see that
the zero-frequency AdS/CFT conductivity is given by
\be \label{condAdS}
\sigma = \frac{1}{g_{d+1}^2}\sqrt{\frac{-g}{g_{rr}g_{tt}}}g^{zz}\biggr|_{r_0} \ .
\ee
One can also derive the above result by writing down an effective action for $A_i$ in the gauge $A_r =0$, in which case one finds an effective action of the form \eqref{ree1} with effective scalar coupling given by
\be \label{eff_scal_coup}
\le({1 \ov q}\ri)_{\rm EM} = {1 \ov g_{d+1}^2} g^{zz}
\ee
From (\ref{ero}), we again find (\ref{condAdS}).  While the DC conductivities for various specific backgrounds have been found in the literature (see e.g.~\cite{herzog,dilepton,centcond}), the general formula \eqref{condAdS} appears to be new. 

Now let us specialize to $d=3$, (2+1)-dimensional field theories, in which case
the metric dependence in \eqref{condAdS} completely cancels and we find
\begin{equation} \label{2+1d}
\sigma = \frac{1}{g_4^2(r_0)} \ .
\end{equation}
This formula was previously derived  for $(2+1)$-dimensional CFTs (in which $g_4$ is necessarily constant)~\cite{herzog,centcond}. We have now shown that it applies to any theory (conformal or not) with a gravity dual. This formula displays a sort of ``bulk universality'' in that it appears very general from the gravity point of view but (unlike $\eta/s$) does not lead to any universality predictions from the field theory perspective, in this case because the bulk gauge coupling $g_{4}^2$ typically does not have a model-independent dual interpretation. We note however that if we restrict attention to CFTs, then $g_{4}^2$ \emph{can} be related to the two-point function of the current at zero temperature, which has the form
\be \label{twze}
\langle \sJ_\mu(x) \sJ_\nu(0) \rangle_{\mathrm{CFT}} = \frac{k}{x^4}\left[\delta_{\mu\nu} - 2\frac{x_\mu x_\nu}{x^2}\right] \
\ee
with $k$ given by $k = \frac{2}{\pi^2 g_{4}^2}$~\cite{freedman}. We thus have the following general expression for the conductivity of any CFT$_3$ with a gravity dual\footnote{In writing down both~\eqref{2+1d} and $k = \frac{2}{\pi^2 g_{4}^2}$ we have assumed a specific normalization for the boundary current $\sJ^\mu$. But Equation~\eqref{uncon} is normalization independent.}
\be \label{uncon}
\sigma = \frac{\pi^2 k}{2} \ .
\ee
The existence of such a relation is nontrivial. For related discussion see \cite{centcond,herzog}.

For completeness, we also note that in $d=3$, one can add to the Lagrangian (\ref{Max})
a theta term given by,
\be
{1 \ov 4} \int d^{4} x \, \th (r) \, \ep^{MNPQ} \, F_{MN} F_{PQ} \ ,
\ee
where we have allowed the $\theta$ parameter to depend on $r$. With this addition, the canonical momentum with respect to a $r$-foliation for $A_\mu$ becomes
\be
j^\mu =  -\frac{1}{g_{4}^2 (r)}\sqrt{-g}F^{r \mu} + {\th (r) \ov 2} \ep^{\mu \nu \lam} F_{\nu \lam} \ .
\ee
We immediately see that now $\sig^{ij}$ has an off-diagonal component (Hall conductivity) given by
\be \label{hacon}
\sig^{12} = - \sig^{21} = \th(r \to \infty) \ .
\ee
This result is exact and (unlike the corresponding result for the diagonal conductivity) does not depend on conditions of horizon regularity or the low frequency limit. Note that
if $\th(r)$ is a nontrivial function~\eqref{hacon} differs from the horizon response, which is $(\sigma_{\mathrm{mb}})^{12} = \theta(r_0)$.

For other dimensions the metric-dependence in (\ref{condAdS}) no longer cancels. Nevertheless we find that the dependence on the metric is essentially dictated by dimensional analysis, and one can still write the conductivity in a form similar to (\ref{2+1d}). We first rewrite (\ref{condAdS}) as
\be \label{cods}
\sig = {1 \ov g_{d+1}^2} (g^{zz})^{d-3 \ov 2} \bigg|_{r=r_0}
\ee
Now recall that in $d$-dimensions $\sig$ has mass dimension $d-3$ (so does $1/g_{d+1}^2$) and
$(g^{zz} (r_0))^\ha$ is the conversion factor between a boundary length scale and the corresponding proper length at the horizon.
Now suppose $l$ is a characteristic scale in the boundary theory (e.g. the inverse temperature), then
$l_{\rm hor} = (g^{zz})^\ha l $ is the corresponding proper length scale at the horizon.
We can now write (\ref{cods}) as
\be \label{conO}
\tilde \sig = {1 \ov \tilde g_{d+1}^2 (r_0)} , \qquad \tilde \sig = \sig l^{d-3}, \quad
{1 \ov \tilde g_{d+1}^2 (r_0)} = {l_{\rm hor}^{d-3} \ov g_{d+1}^2 (r_0)} \ .
\ee
We have constructed the dimensionless conductivity $\tilde{\sigma}$ by rescaling $\sigma$ by a boundary length scale $l$ and similarly constructed the dimensionless horizon gauge coupling $\tilde{g}_{d+1} (r_0)$ by rescaling $g_{d+1} (r_0) $ by the corresponding horizon length scale $l_{\rm hor}$. Equation (\ref{conO}) now again shows ``bulk universality'' with a dimensionless conductivity solely determined by a dimensionless
effective coupling constant. The non-universal metric dependence in (\ref{cods}) is hidden in the conversion factor between the length scales of the boundary and horizon.

As a concrete example, let us look
at a CFT with a gravity dual given by $AdS_{d+1}$ with $d \neq 3$, in which a black hole (with flat section) has the metric
\begin{equation}
ds^2 = \frac{r^2}{R^2}\left(-f(r)dt^2 + d\vec{x}_p^2\right) + \frac{R^2}{f(r)r^2} dr^2
\end{equation}
with $f(r) = 1-\left(r_0/r\right)^{d}$. The Hawking temperature is given by
$T = {d r_0 \ov 4 \pi R^2}$. Applying (\ref{cods}) we find (as shown earlier in \cite{centcond}) that
\begin{equation}
\sigma_{\mathrm{CFT_d}} = \frac{1}{g_{d+1}^2(r_0)}\left(\frac{r_0}{R}\right)^{d-3} \ .
\end{equation}
The corresponding dimensionless conductivity and horizon coupling constant are
$\tilde \sig_{\mathrm{CFT_d}} = \sig_{\mathrm{CFT_d}}  T^{3-d}$
and ${1 \ov \tilde g_{d+1}^2 (r_0)} =  \le({4 \pi \ov d}\ri)^{d-3} {R^{d-3} \ov g_{d+1}^2 (r_0)}$ respectively. Note that it is natural to define the dimensionless coupling constant at the horizon
by normalizing it with the AdS curvature scale. The dimension-dependent prefactor
$\le({4 \pi \ov d}\ri)^{d-3}$ reflects the dimension-dependence of the conversion
from a boundary length scale to that at the horizon.

To conclude this section, we see that a generic transport coefficient of an operator $\sO$ dual to a massless mode in the bulk can be expressed in terms of geometric quantities at the horizon. We emphasize that our analysis depends critically on the existence of a non-degenerate horizon; e.g. for extremal black holes our results do not apply and the behavior of transport coefficients can be quite different. While here we have only discussed the shear viscosity and DC conductivity explicitly, there could be many other applications including e.g., the calculation of the quark diffusion coefficients $\kappa_T$ and $\kappa_L$
discussed in~\cite{CasalderreySolana:2006rq}. The understanding developed here should also make it easier to search for other transport coefficients that exhibit universality.

\section{Flow From the Horizon to the boundary} \label{sec:4}


In the previous section we saw that in the low frequency limit the response of the
boundary fluid is precisely captured by that of the horizon fluid in the membrane paradigm. This happened because the evolution of $\Pi$ and $\p_t \phi$ (or $j^i$ and $F_{ti}$ for a vector field) along the radial direction from the horizon to the boundary
is trivial, and thus the AdS/CFT response depends only on the structure of the horizon. Note nonetheless that the natural definitions of the classical membrane paradigm currents $\Pi_\mathrm{mb}$ and $J^{\mu}_\mathrm{mb}$ differ from their AdS/CFT counterparts $\Pi$ and $j^\mu$ by factors of $\sqrt{-\gamma}$, different index placements, etc. The differences are physically relevant; for example for $d \neq 3$, the boundary conductivity contains extra  temperature-dependence, which should be contrasted with the perspective of a local observer at the horizon membrane who always observes that the conductivity is given by the inverse of the local coupling constant.

The membrane shear viscosity over entropy density ratio (\ref{memD}) does agree exactly with that of the boundary theory since $\eta_{\mathrm{mb}}$ and $s_{\rm mb}$ differ from the corresponding boundary theory quantity by a common factor which cancels in the ratio.
This relation has been noted before \cite{damour,starmem} and the connection between this and the AdS/CFT result is now clear. As is discussed in the conclusion to this paper, the connection between the membrane paradigm bulk viscosity and that of the dual theory is more subtle.


Away from the low frequency limit, the evolution~\eqref{eom1}--\eqref{eom2} from the horizon to the boundary becomes nontrivial and depends on the full geometry. Indeed, the finite frequency/momentum response of an actual strongly coupled quantum field theory is expected to display complicated structure which simply does not exist in the frequency-independent classical membrane paradigm and should arise as we move outwards from the horizon.

To see this more explicitly, we consider a fictitious membrane at each constant radius along the radial direction and introduce a linear response function for each of them,
\be
\bar{\chi} (r; k_\mu) = {\Pi (r, k_\mu) \ov  i \om \phi (r, k_\mu)} \ .
\ee
Note the above expression is defined for all $r$ and $k_\mu$. At $r \to \infty$, $\bar{\chi} (r \to \infty; k_\mu) = -{G_R (k_\mu) \ov i \om}$, which is the finite momentum boundary response
function, and at the horizon $r \to r_0$, $\bar{\chi} (r \to r_0; k_\mu) = \chi$, as
a result of~\eqref{pibc}--\eqref{ero}.

From (\ref{eom1}) and \eqref{eom2}, one can derive a flow equation for $\bar{\chi} (r; k_\mu)$ which governs its  evolution from the horizon to the boundary
\begin{equation}
\p_r \bar{\chi}(r) = i\omega\sqrt{\frac{g_{rr}}{g_{tt}}}\left[\frac{\bar{\chi}^2}{\Sigma_\phi(r)} - \Sigma_\phi(r) \left(1 -
\frac{k^2}{\omega^2}\frac{g^{zz}}{g^{tt}}\right) \right]
\label{eprp}
\end{equation}
where
\be \label{flow_chi}
\Sig_\phi (r) =  {1 \ov q(r)} \sqrt{\frac{-g} {g_{rr}g_{tt}}}, \quad k^2 = \vec k \cdot \vec k \ .
\ee
The above equation makes manifest that as $k \to 0$ and $\om \to 0$, $\bar{\chi}$ is independent of $r$.
Furthermore as $r \to r_0$ in order for the solution to be regular the bracketed quantity in \eqref{eprp} must vanish, from which we recover
\be \label{inee}
\bar{\chi} (r_0, k_\mu) = \Sig_\phi (r_0) = \chi
\ee
where in the second equality we have used \eqref{ero}. It is important to emphasize that the full momentum response at the horizon $\bar{\chi}(r_0,k_\mu)$ automatically corresponds to only to the zero momentum limit of the boundary response, i.e. $\bar{\chi}(r_0,k_\mu) = \bar{\chi}(r \to \infty, k_\mu \to 0)$.

With ``initial condition'' \eqref{inee} the flow equation \eqref{eprp}
can then be integrated from the horizon to infinity to obtain the AdS/CFT response for all $\omega, k$.
The evolution to infinity represents a gradual incorporation of higher-momentum modes. It would be nice to find a precise connection between the flow equation~\eqref{eprp} and the boundary RG flow, possibly in the framework of an exact RG equation~\cite{Polchinski:1983gv}. Note that in general the solution to \eqref{eprp} will involve divergences as $r \to \infty$; these are the familiar UV divergences of the field theory and can be removed by the usual procedure of holographic renormalization.  

We now turn to a $U(1)$ vector field. Without loss of generality, we can take the momentum to be along the $z$ direction.
The conductivities introduced in \eqref{defsigma} and \eqref{Asigma} then naturally separate into two groups, the longitudinal conductivity $\sig_L (k_\mu)
= \sig^{zz} (k_\mu)$  along $z$ direction, and the transverse conductivity  $\sig_T (k_\mu)
= \sig^{xx} (k_\mu)$ along spatial directions not $z$, i.e. $(x,y,...)$.
As for the scalar case, this motivates us to introduce
longitudinal and transverse  $r$-dependent ``conductivities'' respectively,
\be
\bar{\sigma}_L (r; k_\mu) = {j^z(r, k_\mu) \ov F_{zt}(r,k_\mu)}, \qquad
\bar{\sigma}_T (r; k_\mu) = {j^x(r, k_\mu) \ov F_{xt}(r,k_\mu)} \ .
\label{def_sigma_r}
\ee
$\bar{\sigma}_{L,T} (r\rightarrow\infty; k_\mu)$ are the boundary theory responses, and are related to the full retarded correlator $G_R^{\mu\nu}$
of $\mathcal{J}_{\mu}$ as in \eqref{Asigma}.

The Maxwell equations for the vector field in the bulk similarly separate into two groups:
a ``longitudinal'' channel involving fluctuations along $(t,z)$ and a ``transverse'' channel involving fluctuations along all other spatial directions. Using the Maxwell equations in the respective channel we can then derive the flow equation for $\bar \sig_{L,T}$.
Defining
\begin{equation}  \label{def_big_Sigma}
\Sigma_A(r) = \frac{1}{g_{d+1}^2}\sqrt{\frac{-g}{g_{rr}g_{tt}}}g^{zz}
\end{equation}
we show in Appendix \ref{app:A} that $\bar \sig_L$ satisfies an equation of the form
\begin{equation}
\partial_r\bar{\sigma}_L = i\omega\sqrt{\frac{g_{rr}}{g_{tt}}}\left[\frac{\bar{\sigma}_L^2}{\Sigma_A(r)}\left(1 -
\frac{k^2}{\omega^2}\frac{g^{zz}}{g^{tt}}\right) - \Sigma_A(r)\right] \ .
\label{sigmaflow}
\end{equation}
$\bar{\sig}_T$ satisfies the same equation as \eqref{eprp}, but with $\Sig_\phi$ replaced by $\Sig_A$. Regularity at the horizon membrane again provides the initial data at the horizon
\be
\bar{\sigma}_{T,L}(r_0, k_\mu) = \Sigma_A(r_0) = \sig  \ .
\ee

There is a curious relation between the flow equations for $\bar \sig_L$ and
$\bar \sig_T$. It can be readily checked from \eqref{sigmaflow} and \eqref{eprp} that the quantity ${1 \ov \bar \sig_L}$ satisfies an equation which is identical to that of $\bar \sig_T$ after
the replacement $\Sig_A (r) \to {1 \ov \Sig_A (r)}$. This can be interpreted as a relation between the conductivities of two different theories, and is discussed in Appendix \ref{app:dual}.

\section{Diffusion at the boundary and at the horizon} \label{sec:diff}

In section \ref{sec:3} we showed that in the low frequency limit the linear response at the boundary is fully captured by the response of the horizon. In section \ref{sec:4} we showed that finite frequency/momentum response of the boundary cannot be captured by that at the horizon since the horizon response always corresponds to the low frequency limit of the boundary theory. In this section we examine linear response at the next order in the derivative expansion using the examples of diffusion of charge and
momentum density. The discussion serves to highlight the differences between the diffusion at the horizon and the boundary. It also provides a straightforward application of the flow equations derived in the last section, which we use to give a simple derivation of the diffusion constant.

\subsection{Charge diffusion}

Consider disturbing the thermal equilibrium of the boundary theory by a small nonuniform perturbation of charge density varying along the $z$ direction. The charge gradient generates a nonvanishing current $\sJ^z$ and eventually the charge diffuses away back into thermal equilibrium. To lowest order in the derivative expansion the diffusion process is governed by the dispersion relation
\be
\om = - i D  k^2, \qquad k = k_z
\label{diee}
\ee
where $D$ is the diffusion constant. In the linear response regime \eqref{defsigma}-\eqref{Asigma}, equation \eqref{diee} appears as a pole in the retarded Green function $G^{zz}_R$, since there is a nonzero current $\sJ^z$ even in the absence of an external field.

To study the diffusion process from gravity  we should thus examine the longitudinal channel in the regime $\omega \sim k^2$ and $\om /T \ll 1, k/T \ll 1$. Assuming this scaling and taking $\bar{\sigma}_L \sim O(1)$, we obtain from \eqref{sigmaflow}
\be
{\partial_r\bar{\sigma}_L \ov \bar{\sigma}_L^2} = -\frac{i k^2}{\omega}\frac{\sqrt{g_{rr}g_{tt}}}{\Sigma_A g_{zz}} \ .
\ee
The solution to this equation with initial condition given by \eqref{horizcond} is
\be
\frac{1}{\bar{\sigma}_L(r)} = \frac{1}{\sig} + i\frac{k^2}{\omega}\int_{r_0}^r dr' {\sqrt{g_{rr}g_{tt}}g^{zz} \ov \Sigma_A}
\ee
where we have used that the DC conductivity $\sig$ is given by $\sig = \Sigma_A(r_0)$.
Now recalling that $\sig^{zz} (k_\mu) = \sig_L (k_\mu) = \bar{\sigma}_L(r\to \infty, k_\mu)$ and using \eqref{Asigma}, we find that
\be
G_R^{zz}(k_\mu) = \frac{\omega^2 \sig}{i\omega - Dk^2}
\label{corr_form}
\ee
where the diffusion constant $D$ is given by the integral  
\be
D = \sig
\int_{r_0}^\infty dr' \frac{g_{rr}g_{tt}}{\sqrt{-g}}g_{d+1}^2 \ .
\label{diff_const}
\ee
Equation \eqref{diff_const} is equivalent to the diffusion constant derived in \cite{kss_stretch,starmem} once we substitute the explicit expression~\eqref{condAdS} for $\sig$.

Note that the diffusion constant $D$~\eqref{diff_const} cannot be generically written 
in terms of horizon quantities. 
Now using the Einstein relation~\footnote{Note that with the Einstein relation below \eqref{corr_form} precisely has the form which one expects
from  hydrodynamics
$$
G^R_{zz} = \frac{\omega^2 \Xi D }{i\omega - Dk^2} \ .
$$}
\be
\Xi D  = \sig
\ee
where $\Xi$ is the charge susceptibility, we find a general expression for
$\Xi$,
\be \label{Xi}
\Xi =  \left[\int_{r_0}^\infty dr' \frac{g_{rr}g_{tt}}{\sqrt{-g}}g_{d+1}^2\right]^{-1} \ .
\ee
In Appendix~\ref{app:C} we give an alternative derivation of \eqref{Xi} for an arbitrary charged black brane; thus the logic of this section can also be viewed as a proof that any black brane obeys the Einstein relation.

Having derived the retarded Green function \eqref{corr_form} in the diffusion regime,
let us now find bulk solutions to the Maxwell equations that are dual to these diffusive modes in the gauge theory. This in particular will enable us to compare
the explicit diffusion processes on the horizon membrane and at the boundary.
For this purpose we again examine the longitudinal channel Maxwell equations \eqref{flow2} and \eqref{bianchi} in the limit of small $\om$ and $k$ but finite $\omega \sim k^2$,
\bea
\partial_r j^z & = & \sO(\omega F_{zt}),  \\
\partial_r F_{zt} & = & \frac{ik^2}{\omega} \frac{g_{rr}g_{tt} g_{d+1}^2}{\sqrt{-g}}j^z + \sO(\omega j^z) \ .
\eea
The above equations can be immediately integrated to give
\bea
j^z(r) & = & j^z(r_0) = \mathrm{const} \label{diff1},  \\
F_{zt}(r) & = & F_{zt}(r_0)\left[1 + \sig \frac{ik^2}{\omega}\int_{r_0}^r dr'\frac{g_{rr}g_{tt}g_{d+1}^2}{\sqrt{-g}}\right] \nonumber \\ \label{diff2}
\eea
where we have used the boundary condition $j^{z}(r_0) = \sig F_{zt}(r_0)$ at the horizon.

Since we are interested in a diffusion process 
we should be looking for gauge theory configurations where the applied electric field is zero. Requring $F_{zt}(r \to \infty) = 0$  enforces a relation between $\omega$ and $k$ that is exactly the dispersion relation~\eqref{diee} with $D$ given by \eqref{diff_const}. When this relation is satisfied, in the dual picture we see standard diffusion with no electric field and $\langle \sJ^z \rangle_{\mathrm{QFT}}$ decaying with time.

This should be contrasted with the behavior observed at the horizon. As $j^z$ is constant throughout the spacetime, an observer at the horizon membrane will also see a
current evolving with precisely the same behavior as that at the boundary. Indeed, diffusive behavior on stretched horizons has been noted before~\cite{kss_stretch,saremi}. 
However, the electric field $F_{zt}$ at the horizon is \emph{not} zero; indeed, we know that at the horizon the electric field exactly tracks the current via the boundary condition \eqref{horizcond}:
\be
j^{i}(r_0) =  \sig F_{it}(r_0) \ ,
\label{horizcond2}
\ee
and thus cannot be zero. Thus a local observer at the horizon has a rather different interpretation; he sees a nonzero electric field that is decaying with time as it falls into the horizon, and this electric field directly induces a membrane current via the simple response \eqref{horizcond2}. As emphasized before, \eqref{horizcond2} in fact exists for all $\omega,k$; however if $\omega, k$ satisfy the diffusion relation~\eqref{diee} with \eqref{diff_const} then the electric field perturbations will vanish at infinity and the boundary interpretation of this configuration will be diffusion.

\subsection{Momentum diffusion}

For momentum diffusion one considers in the boundary theory a small nonuniform perturbation in momentum density $\delta \sT_{at}$ varying along the $z$ direction, where $a$ is any spatial direction $x$,$y$, etc. not equal to $z$. The diffusion current is given by $\delta \sT_{az}$ and the diffusion constant $D_s$ can be written in terms of boundary quantities  as~(for a recent review see~\cite{viscreview})
\be
D_s = {\eta \ov \ep + p}
\ee
where $\ep$ and $p$ and energy and momentum density.

The gravity modes corresponding to these components are $h_{at}, h_{az}$
which decouple from the other components in the gauge $h_{ar} =0$. The corresponding
bulk canonical momenta are $T^{a \mu}$ given by \eqref{stens}, now with $\mu = (z,t)$.
As pointed out in \cite{kss_stretch}, the quickest route to the bulk equations of motion in this channel is via a Kaluza-Klein reduction in the $a$ direction. In this case the relevant modes $h^{a}_\mu$ ($\mu = (t,z)$) can be considered components of a gauge field $A_\mu$ with a metric-dependent gauge coupling. The precise mapping is
\be
h^{a}_{\mu} = A_\mu, \qquad T^\mu_a = j^{\mu}, \qquad \frac{1}{g_{d+1}^2(r)} = \frac{g_{xx}(r)}{16\pi G_N} \
\ee
whose derivation we review in Appendix~\ref{app:B}.
We can now immediately take over all of the results of the previous section: e.g.
the corresponding ``DC conductivity'' is given by the entropy density ${s \ov 4 \pi}$ and
the retarded correlator of $\sT_{a}^{z}$ with itself is given by the analog of \eqref{corr_form}
\be
G_{R}^{az,az} = \frac{s}{4\pi}\frac{\omega^2}{i\omega - D_s k^2}
\ee
with the diffusion constant $D_s$ given by
\be
D_s = {\eta \ov \ep + p} = 
4 G_N \, s \,  \int_{r_0}^\infty dr' \frac{g_{rr}g_{tt}}{\sqrt{-g}g_{xx}}
\label{sm_diffconst}
\ee
This equation is again equivalent to that  in \cite{kss_stretch}. From \eqref{sm_diffconst} and \eqref{etas} we now derive a general formula for $\ep + p$,
\be
{1 \ov \ep + p} = 16 \pi G_N \, \int_{r_0}^\infty dr' \frac{g_{rr}g_{tt}}{\sqrt{-g}g_{xx}} \ .
\ee
Note that the analog of the ``electric field'' is a gauge-invariant \footnote{The gauge transforms in this section are diffeomorphisms of the $(r,t,z)$ sector; the effect of these on the $h^a_\mu$ perturbations is that of the $U(1)$ of the effective gauge field $A_\mu$.} combination $\mathcal{E}_z$ of metric coefficients
\be
\mathcal{E}_z \equiv \partial_z(h^a_t) - \partial_t(h^a_z) \ .
\ee

We will not repeat here the analysis of last subsection. Everything can be carried over with a change of notation as described above.
We do emphasize again that the diffusion process
at the boundary is different from that on the horizon.
An observer at the horizon will see that the horizon metric is deformed; this deformation is directly inducing a fluid flow on the membrane via the response $T^z_a (r_0) = {s \ov 4 \pi} \mathcal{E}_z$.

Finally, note that diffusive behavior is not specific to the examples considered. Given a general pair $(\Pi, \partial_t\phi)$, the essential ingredient in this analysis is the existence of a scaling limit for $\omega$, $k$ where the response $\Pi$ is constant in $r$ but the source $\partial_t\phi$ is not (see \eqref{diff1}, \eqref{diff2}). This can be translated into a constraint on general effective actions such as \eqref{moreG}: we would like a limit where $P(r;k_\mu)/\omega \to 0$ but $Q(r;\omega,k)\omega \to O(1)$. This allows us to construct a bulk solution where the source $\phi(r \to \infty)$ vanishes but $\Pi(r\to\infty)$ does not. 

\section{Conclusion and discussion} \label{sec:concl}

We have shown that there is a precise sense in which the long-wavelength limit of a boundary  theory at finite temperature
is determined  by the horizon geometry of its gravity dual. We derived expressions for various transport coefficients in
terms of components of the metric evaluated at the horizon; this sheds light on the origin of the universality of the shear viscosity
and resulted in a general formula for the conductivity \eqref{condAdS}. 
At finite frequency/momentum, however, propagating the information from the horizon to the boundary  requires solving a nontrivial flow equation which describes how the full AdS geometry encodes the higher momentum degrees of freedom of the boundary theory.
The examples of charge and momentum diffusion provide illustrations of this flow in a very simple context. 

Note that the relation between the membrane paradigm bulk viscosity (which is negative) and the bulk viscosity of a conformal fluid (which is exactly zero) is more subtle. Here the relevant degree of freedom $h_i^i$ does not satisfy a simple equation like $h_x^y$. Instead, it enters in a nontrivial way into the Hamiltonian constraint of general relativity in the bulk and thus (in the absence of a background scalar profile) is not actually a propagating degree of freedom, leading to a vanishing bulk viscosity for conformal theories. If one turns on a nontrivial massive scalar background (corresponding to a deviation from conformality) then fluctuations of this field can mix with the graviton $h^i_i$, bringing it to life and allowing a nonzero bulk viscosity\footnote{For explicit calculations see \cite{gubser,bulk1}.}. Thus a systematic study of bulk viscosity will involve fluctuations of massive fields. This will dramatically change the structure of flow equations such as \eqref{eprp}; in particular, the flow will no longer be trivial even in the low frequency limit and it is likely that the membrane response will not adequately capture the low-frequency AdS/CFT response. We defer such complications to a future publication.

We close by noting that the classical membrane paradigm fluid can be seen to play a satisfying new role in the holographic description of a strongly coupled field theory at finite temperature. We hope that it may be a practically useful role as well, both for identifying what quantities are expected to display universal behavior and as a technical tool for simplifying future hydrodynamic
computations in gauge-gravity duality.

\begin{acknowledgments}

We would like to thank D.~Anninos, C.~Athanasiou, A.~Buchel, J.~Casalderrey-Solana, T.~Faulkner, P.~Kovtun, D.~Mateos, J.~McGreevy, P.~Petrov, K.~Rajagopal, D.~Son, A.~Starinets, and U.~Wiedemann  for helpful discussions.
Research supported in part by
the DOE
under
contracts
\#DF-FC02-94ER40818.
HL is also supported
in part by the A.~P.~Sloan Foundation and the 
DOE OJI program.  NI is supported in part by National Science Foundation (NSF)
Graduate Fellowship 2006036498. 

\end{acknowledgments}

\begin{appendix}
\section{In-falling boundary conditions and horizon regularity} \label{sec:inf}
Here we demonstrate using the example of a scalar field that the in-falling boundary condition used in standard AdS/CFT calculations is equivalent to the condition of horizon regularity used in the membrane paradigm. At the horizon $r \to r_0$, the metric may be written
\be
g_{tt} = c_0 (r-r_0), \qquad g_{rr} = {c_r \ov r-r_0}
\ee
One finds that near the horizon the equation for $\phi$ is given by
\be
\sqrt{c_0 \ov c_r} (r-r_0) \p_r \le(\sqrt{c_0 \ov c_r} (r-r_0) \p_r \phi \ri)
+ \om^2 \phi =0
\ee
which gives
\be \label{torN}
\phi \propto e^{ -i \om (t \pm x)}, \qquad dx = \sqrt{g_{rr} \ov g_{tt}} dr 
\ee
The in-falling boundary condition implies that we should take the positive sign in the exponent. In that case it is clear that the solution can be written only in terms of the Eddington-Finkelstein coordinate $v$ defined in \eqref{EFC}:
\be
\phi \propto e^{-i\om v}, \qquad dv = dt + \sqrt{g_{rr} \ov g_{tt}} dr
\ee
The fact that $\phi$ depends only on the nonsingular coordinate $v$ at the horizon is precisely the condition of horizon regularity used in the membrane paradigm. Note if we integrate the definition of $v$ in a small neighborhood of the horizon and use the formula for the inverse Hawking temperature $\beta = 4\pi\sqrt{\frac{c_r}{c_0}}$ we obtain
\be v = t + \frac{\beta}{4\pi} \ln(r-r_0) \to \phi \propto (r-r_0)^{-\frac{i\omega\beta}{4\pi}} e^{-i\omega t}
\ee
which is recognizable as the standard form of the in-falling boundary condition. 

\section{Bulk Maxwell equations and flow equations for conductivities} \label{app:A}

Here we assemble the relevant components of the bulk Maxwell equations, given by the variation of the action
\be \label{appmaxaction}
S = -\int d^{d+1}x \sqrt{-g} \frac{1}{4 g_{d+1}^2(r)} F_{MN} F^{MN}
\ee
As before $j^{\mu}$ is the momentum conjugate to $A_\mu$ with respect to a foliation by constant-$r$ slices \eqref{def_j}:
\be
j^{\mu} = -\frac{1}{g_{d+1}^2}\sqrt{-g}F^{r \mu}
\ee
Rather than working with the gauge potentials $A_M$, we will write all equations in terms of gauge-invariant objects such as $F_{\mu\nu}$ and $j^\mu$; this involves the manipulation of a larger number of equations but makes the physical interpretation more transparent. We will also assume that the background configuration of the gauge field is trivial; in principle fluctuations around a nontrivial gauge field background can couple to the metric or other fields and require a more detailed analysis. Note that in \eqref{appmaxaction} $g_{d+1}^2(r)$ can be given by the background value of a nontrivial scalar field, as symmetry arguments guarantee that fluctuations of such a scalar will decouple from the Maxwell perturbations.

If we take the momentum to be along the $z$ direction,
these equations naturally separate into two groups; a
``longitudinal'' channel involving fluctuations along $(t,z)$ and a ``transverse'' channel involving fluctuations along all spatial
directions not $z$, i.e. $(x,y,...)$. Defining $G = \sqrt{-g}/g_{d+1}^2$ for notational convenience, we find that the longitudinal channel
is governed by two dynamical equations:
\begin{align}
-\partial_r j^t - G g^{tt}g^{zz}\partial_z F_{zt} = 0 \label{flow1}\\
-\partial_r j^z + G g^{tt}g^{zz}\partial_t F_{zt} = 0 \label{flow2}
\end{align}
as well as the conservation of $j^{\mu}$ and the Bianchi identity:
\begin{align}
\partial_t j^{t} + \partial_z j^{z} = 0 \label{jconservation} \\
-\frac{g_{rr}g_{zz}}{G}\partial_t j^{z} - \frac{g_{rr}g_{tt}}{G}\partial_z j^{t} + \partial_r F_{zt} = 0 \label{bianchi}
\end{align}
We would like to derive a flow equation for $\bar{\sigma}_L(r;k_\mu) \equiv j^{z}/F_{zt}$. We begin by taking a single derivative:
\begin{equation}
\partial_r \bar{\sigma}_L = \frac{\partial_r j^{z}}{F_{zt}} - \frac{j^{z}}{F_{zt}^2}\partial_r F_{zt}
\end{equation}
We now use attack the right-hand side, using \eqref{jconservation} to eliminate $j^{t}$ in favor of $j^{z}$, \eqref{bianchi} to eliminate
$\partial_r F_{zt}$ in favor of $j^{z}$, and \eqref{flow1} to eliminate $\partial_r j^{z}$ in favor of $F_{zt}$. The final
differential equation for $\bar{\sigma}_L$ is
\begin{equation} \label{sigmaLflow_app}
\partial_r\bar{\sigma}_L = i\omega\sqrt{\frac{g_{rr}}{g_{tt}}}\left[\frac{\bar{\sigma}_L^2}{\Sigma_A(r)}\left(1 -
\frac{k^2}{\omega^2}\frac{g^{zz}}{g^{tt}}\right) - \Sigma_A(r)\right],
\end{equation}
where as in the text \eqref{def_big_Sigma} we have defined $\Sigma_A(r) = \frac{1}{g_{d+1}^2}\sqrt{\frac{-g}{g_{rr}g_{tt}}}g^{zz}$.

Similarly, the transverse channel is governed by a dynamical equation and two constraints from the Bianchi identity.
\begin{align}
-\partial_r j^{y} - G g^{tt}g^{yy}\partial_t F_{ty} + G g^{zz}g^{yy} \partial_z F_{zy} = 0 \label{flowT1}\\
\partial_r F_{yt} - \frac{g_{rr} g_{yy}}{G}\partial_t j^y = 0 \label{flowT2} \\
\partial_z F_{ty} + \partial_t F_{yz} = 0 \label{bianchiT}
\end{align}
To find the flow equation for $\bar{\sigma}_T \equiv j^y/F_{yt}$ we follow a procedure directly analogous to that above, using \eqref{bianchiT} to eliminate $F_{yz}$ in favor of $F_{yt}$, \eqref{flowT1} to
eliminate $\partial_r j^y$ in favor of $F_{yt}$, and $\eqref{flowT2}$ to eliminate $\partial_r F_{yt}$ in favor of $j^y$. The resulting equation is
\begin{equation}
\partial_r\bar{\sigma}_T = i\omega\sqrt{\frac{g_{rr}}{g_{tt}}}\left[\frac{\bar{\sigma}_T^2}{\Sigma_A(r)} - \Sigma_A(r)\left(1 - \frac{k^2}{\omega^2}\frac{g^{zz}}{g^{tt}}\right)\right] \ .
\label{sigmaTflow}
\end{equation}

\section{A curious relation and electric-magnetic duality} \label{app:dual}
Examination of \eqref{sigmaLflow_app} and \eqref{sigmaTflow} shows that the quantity ${1 \ov \bar \sig_L}$ satisfies an equation which is identical to that of $\bar \sig_T$ after
the replacement $\Sig_A (r) \to {1 \ov \Sig_A (r)}$. Following analogous discussion to that around \eqref{cods} and \eqref{conO}, one can also interpret $\Sig_A (r)$ as an ``effective dimensionless coupling constant'' at each $r$-hypersurface. In other words, if we have two bulk theories $1,2$ whose dimensionless couplings $\Sigma_A^i$ are related by
\be \label{dualSig}
\Sigma_{A}^1(r) = {1 \ov \Sigma_A^2(r)}
\ee
then the respective conductivities of these two theories are related by
\be \label{duality}
\sigma_T^1(k_\mu) = \frac{1}{\sigma_L^2(k_\mu)} \qquad \sigma_L^1(k_\mu) = \frac{1}{\sigma_T^2(k_\mu)}
\ee
This is a peculiar relation which we now explain, starting with the special case when $d = 3$ and $g_4^2$ is constant. Here $\Sig_A = 1/g_4^2$ is independent of position and the inversion $\Sig_A (r) \to 1/\Sig_A (r)$ only involves an inversion of the coupling constant $g_4^2$. At the quadratic level in the supergravity action this coupling can be scaled out and does not affect the dynamics, and so \eqref{duality} actually relates $\sigma_L$ to $\sigma_T$ in the same theory: $g_4^2\sigma_T = 1/(g_4^2\sigma_L)$. This symmetry of the equations of motion is actually a consequence of electric-magnetic duality in four bulk dimensional electromagnetism. Switching $E$ and $B$ is equivalent to interchanging longitudinal electric fields $E_z$ for transverse currents $j^x$ and vice-versa. This relates the two sides of the relations in \eqref{duality}; this connection has been emphasized previously in \cite{herzog}.

To understand why a similar relation might hold in higher dimension, we should note that the bulk dynamics can always be reduced to an effective four-dimensional system by dimensional reduction along all dimensions not equal to ${r,t,x,z}$. The kinetic term for the gauge field is then
\be
S = \int d^{d}x \frac{1}{g_{d+1}^2(r)}\sqrt{-h}{g_{xx}}^{\frac{d-3}{2}}F_{ab}F^{ab},
\ee
where $a,b$ are in the four dimensional space parametrized by ${r,t,x,z}$ and $h_{ab}$ is the metric on this space. The equations of motion from this action are identical to those of four-dimensional electromagnetism with position-dependent gauge coupling
\be
\frac{1}{g_4^2(r)} = \frac{1}{g_{d+1}^2(r)}{g_{xx}}^{\frac{d-3}{2}} = \Sigma_A
\ee
On the other hand this theory is dual by standard four-dimensional electric-magnetic duality to a theory with inverted gauge coupling $g_4^2(r)$. Thus if two $(d+1)$ dimensional bulk theories satisfy \eqref{dualSig}, then the effective four-dimensional dynamics of these two theories are related by electric-magnetic duality. This relation manifests itself in \eqref{duality}. For $d = 3$ and constant $g_{d+1}^2$ such an expression is thought to be related to a non-Abelian generalization of particle-vortex duality  in the boundary theory \cite{herzog}; it would be interesting to find a similar boundary interpretation for the $d \neq 3$ case \eqref{duality}.

\section{The Einstein relation for arbitrary charged black branes} \label{app:C}
In the text we computed independent expressions for $\sigma$ and $D$, the conductivity and diffusion constant for an arbitrary conserved current in any field theory with a gravity dual. To complete our discussion we now compute the charge susceptibility $\Xi$ for an arbitrary charged black brane.

The charge density of the dual theory is $\rho = j^t(r \to \infty)$ and the chemical potential is $\mu = A_t(r \to \infty)$. To evaluate the susceptibility we require $\rho$ to linear order in $\mu$, $\rho(T,\mu) \equiv \Xi(T)\mu$. We consider a static bulk field configuration depending only on $r$.  Examining \eqref{flow1} and \eqref{flow2}, we obtain
\bea
\partial_r j^t & = & 0 \\
\partial_r A_t & = & \frac{g_{rr}g_{tt}}G j^t
\eea
with the immediate solution
\bea
j^t(r) & = &\rho = \mathrm{const} \\
A_t(r) & = &A_t(r_0) + \rho \int_{r_0}^r dr' \frac{g_{rr}g_{tt}}{G}
\eea
Horizon regularity requires $A_t(r_0) = 0$, giving us $\mu = A_t(r \to \infty) = \rho \Xi^{-1}$ with
\be
\Xi = \left[\int_{r_0}^\infty dr' \frac{g_{rr}g_{tt}g_{d+1}^2}{\sqrt{-g}}\right]^{-1}
\ee
Comparing with \eqref{diff_const} and \eqref{condAdS}, we see that the Einstein relation
\be
\sigma = \Xi D
\ee
is indeed satisfied for any black brane.

\section{Dimensional reduction for gravitational shear mode} \label{app:B}

As emphasized in \cite{kss_stretch}, the relevant equations for gravitational shear mode fluctuations can be mapped onto an electromagnetism problem. Consider a metric perturbation of the form
\be \label{metperturb}
g_{a\nu}(r) \to g_{a\nu}(r) + g_{aa}(r)h^{a}_{\nu}(r,t,z)
\ee
where $a$ is a spatial direction that is not equal to $z$. Compare this to the standard form of the metric used in a Kaluza-Klein reduction along the $a$ direction:
\be \label{kkmetric}
ds^2 = g_{MN}dx^M dx^N = g_{\alpha\beta}dx^\alpha dx^\beta + g_{aa}(dx^a + A_\beta dx^\beta)^2
\ee
where the indices $\alpha$, $\beta$ omit the $a$ direction and $A_\beta$ is an effective $(d-1)$ dimensional gauge field. This is exactly the form of the perturbation \eqref{metperturb}, provided we set $A_\beta = h^{a}_{\beta}$. However since nothing depends on the $a$ direction, we can integrate out $a$ from the Einstein-Hilbert action constructed from \eqref{kkmetric}. Standard dimensional reduction formulae (see e.g. \cite{polchinski}) give us the kinetic term for $A$
\be
S = -\frac{1}{16\pi G_N}\int dx^a\int d^{d-1} x\sqrt{-g}g_{aa} F_{\alpha\beta} F^{\alpha\beta}
\ee
where $F$ is the field strength tensor of $A$, which in terms of metric perturbations is $F_{\alpha\beta} = \partial_\alpha h^{a}_\beta - \partial_\beta h^{a}_\alpha$. Here the determinant $\sqrt{-g}$ is that of the full $d$-dimensional metric. This action is of exactly the standard Maxwell form \eqref{appmaxaction} with an effective coupling for the gauge field
\be
\frac{1}{g_{d+1}^2} = \frac{1}{16\pi G_N}g_{xx}
\ee
as claimed in the text.
\end{appendix}

\end{document}